\documentclass[aps,prl,nofootinbib,floatfix,twocolumn,showpacs]{revtex4-1}

\usepackage{pstricks}
\usepackage{graphicx}
\usepackage{bm}
\usepackage{amsmath,amssymb,amsfonts}

\usepackage[sort&compress]{natbib}

\begin{document}

\title{Accelerator experiments check general relativity}

\author{Vahagn Gharibyan}
\email[]{vahagn.gharibyan@desy.de}
\affiliation{Deutsches Elektronen-Synchrotron DESY - D-22603 Hamburg}

\begin{abstract}
The deflection of gamma-rays in Earth's gravitational field is tested 
in laser Compton scattering at high energy accelerators.
Within a formalism connecting the bending angle to the photon's
momentum it follows that detected gamma-ray spectra are inconsistent
with a deflection magnitude of 2.78~nrad, predicted by 
Einstein's gravity theory.
Moreover, preliminary results for 13--28 GeV photons from two different laboratories
show opposite -- away from the beam line -- deflection, amounting to 33.8--0.8~prad. 
These conclusions, however, are applicable only for  
gravitationally sterile high energy electrons.
Much more subtle effects are expected if the gravitational 
deflection of the electrons is taken into account.
\end{abstract}

\pacs{04.80.Cc, 04.20.-q,  29.27.-a}

\maketitle

{\it Introduction.---}
Einstein's general relativity (GR)~\cite{Einstein-GR} is the
currently established theory  of gravitation and has been confirmed
in all observations and experiments to date~\cite{Beringer:1900zz}.
An essential validity check of GR is based on gravitational light bending.
These deflection measurements, which were
started by a spectacular observation of starlight deflection during a solar eclipse
about a century ago~\cite{eddington}, have been expanded to 
radio-waves and have become very precise~\cite{Lebach:1995zz}. 
The most accurate measurements are performed using the gravitational field of the 
Sun~\cite{Shapiro:2004zz}, as the bending increases with  mass.
The results prove that electromagnetic radiation, from radio to visible 
light frequencies, is bent according to GR and follows the curvature of 
space~\cite{Will:2005va}. On a scale of less massive objects, the
light bending strength of the planet Jupiter has also been tested and 
quantified~\cite{Jupiter}. 
For the Earth, however, a check of gravitational bending remains infeasible
because of the smallness of the expected deflection, about 3~nrad (compared with more than 
8~$\mu$rad for the Sun).
The quoted numbers follow from the well-known expression $4GM/c^2R$ for the 
deflection angle~\cite{Landau1975}, when light travels near a mass $M$ with an 
impact parameter $R$ (higher-order terms of deflection are neglected throughout the Letter).
For a light ray grazing the Earth's surface, the total deflection angle is
\begin{equation}
\frac{4 G M_\oplus}{c^2 R_\oplus} \approx 2.78\times 10^{-9},
\label{Eangle}
\end{equation}
where $G$ is the gravitational constant and $c$ is the speed of light.
The bending magnitude for light generated and measured in a laboratory is much 
smaller and is equal to
\begin{equation}
 \frac{2 G M_\oplus}{c^2 R_\oplus}\frac{L}{\sqrt{L^2 + R_\oplus^2}},
\label{Lab-angle}
\end{equation}
where $L$ is the length of light travel~\cite{limited-bending}.
Thus, for a distance of 1m, this angle is only $2\times 10^{-16}$rad and
the light shifts by 0.2 femtometer, which is undetectably small, at least for a 
direct measurement. 
A way to overcome this problem is described in ref.~\cite{slow-light}, 
which is based on the idea of slowing  light down to $v\approx 100$m/s,
in order to substitute $c$ in Eq.(\ref{Lab-angle}) by  $v$ and 
increase the bending magnitude. 

In this Letter, I describe a laboratory method that probes gravitational 
bending of high-energy photon beams.

{\it Gravity as a bending medium.---}
Within the  method I use an idea that gravitation for light
is equivalent to an optical medium. This idea was suggested by  
Einstein and was employed by many authors; see ref.~\cite{de Felice:1971ui} 
and references therein. Felice~\cite{de Felice:1971ui}
has proved that a GR curved space described by  Riemannian geometry 
is identical to the language of classical optics in a flat space medium.
The author, however, has warned that the optical description may be mathematically   
more complicated, although it could be beneficial for solving certain problems.
The deflection of high-energy photons is one such problem.
In a recent paper~\cite{Ye:2007df}, the authors further developed an optical approach 
and suggested the following refractive index for a spherically symmetric 
gravitational field:
\begin{equation}
n=\exp \Bigl(\frac{2 G M}{c^2 R}\Bigr),
\label{refind}
\end{equation}
which, for the Earth's weak field, reduce to 
\begin{equation}
n_\oplus=1+\frac{2 G M_\oplus}{c^2 R_\oplus}.
\label{refearth}
\end{equation}
The latter expression has also been derived by other 
authors~\cite{grav=optics, Boonserm:2004wp, Sen:2010zzf} and 
is equivalent to  Eqs.(\ref{Eangle}) and (\ref{Lab-angle}) 
when one applies  optical tracking with such a refractive index.
The main difference between the gravitational "medium" presented 
by Eq.(\ref{refearth}) and any material medium 
is the bending independence on frequency of light or photon energy,
which is a consequence of the gravity geometrical interpretation or 
the curved space-time concept.  
It will help us to test the gravitational bending since
scattering or interaction angles decrease toward  high energies; 
at some energy, the angle will approach the magnitude of refractivity given in the 
Eq.(\ref{refearth}), interfering with the gravity.  

{\it The Compton process in a gravitational field.---}
A proper process to explore is  high-energy Compton scattering, which is
sensitive to tiny deviations of the refractive index from unity, as described
in ref.~\cite{Gharibyan:2012gp}. 
Using energy-momentum conservation, when at $n\approx 1$ a photon scatters off an 
electron with energy ${\cal E}$, 
the Compton scattering kinematics is given by
\begin{equation}
{\cal E}x - \omega (1+x+\gamma^2 \theta^2) + 
2\omega \Bigl(1 - \frac{\omega}{{\cal E}} \Bigr) \gamma^2 (n-1) = 0,
\label{comp0}
\end{equation}
where \hbox{$x=4\gamma \omega_0\sin^2{(\theta_0/2)}/m$}, 
with $\gamma$ and $m$ being the Lorentz factor and mass of the
initial electron, respectively. The initial photon's energy and angle are denoted
by $\omega_0$ and $\theta_0$, while the refractive index $n$ is in effect for the
scattered photon with energy $\omega$ and angle $\theta$; the angles are defined relative to the initial electron. This kinematic
expression is identical to Eq.(8) from ref.~\cite{Gharibyan:2012gp}  and 
is derived for small refractivity and high energies,  
i.e., the  $\mathcal{O}((n-1)^2)$, $\mathcal{O}(\theta^3)$, and 
$\mathcal{O}(\gamma^{-3})$ terms are neglected.
To determine the outgoing photon's energy, I solve Eq.(\ref{comp0}) for $\omega$ and, 
to leading order of $(n-1)$, I obtain:
\begin{equation}
\omega = \frac{{\cal E}x}{1+x+\gamma^2 \theta^2} \Bigl( 1+ 
\frac{2\gamma^2 (n-1)(1+\gamma^2 \theta^2)}{(1+x+\gamma^2 \theta^2)^2} \Bigr).
\label{comp}
\end{equation}
Writing this formula for the maximal energy
of the scattered photons (Compton edge, at $\theta=0$) in the Earth's gravitational 
field, I obtain:
\begin{equation}
\omega_{max} = \frac{{\cal E}x}{1+x} \Bigl( 1+ 
\frac{2\gamma^2 (n_\oplus -1)}{(1+x)^2}   
\Bigr),
\label{wmax}
\end{equation}
where the Earth's light bending refractivity $n_\oplus -1$ is amplified by $\gamma^2$, 
allowing one measure it by detecting the extremal energy of the scattered photons 
$\omega_{max}$, or electrons ${\cal E}-\omega_{max}$.  
\begin{figure}[b]
\centering
\includegraphics[scale=0.47]{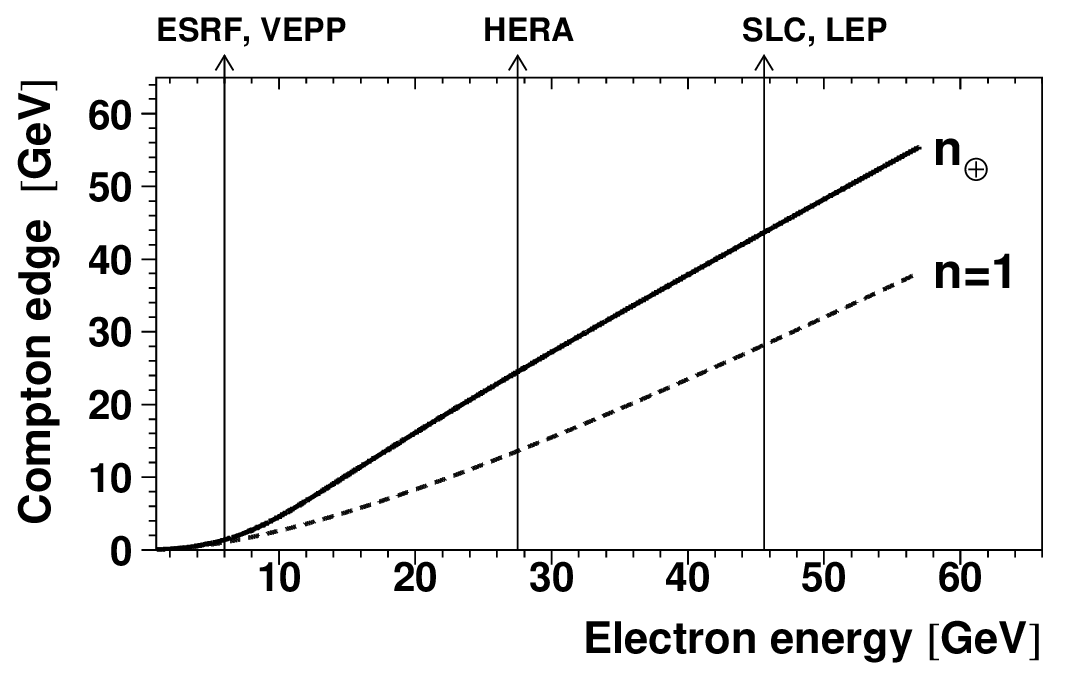}
\caption{\label{fig1}
Compton scattered photons' maximal energy (Compton edge)
dependence on the initial electron energy 
for a head-on collision with 532nm laser light.
Solid and dotted lines correspond to the refractive index of the 
gravitational field at the Earth's surface ($n_\oplus = 1+1.39\times 10^{-9})$, 
and free space (n=1) respectively.}
\end{figure}
In order to estimate the method's sensitivity, I calculate the Compton edge
for an incident photon energy 2.32~eV (the widely popular green laser) 
at different energies of the accelerator electrons.  
The resulting dependencies for a free space ($n=1$) and the Earth's gravity-induced 
refractivity  are presented in Fig.~\ref{fig1}.
The plot shows considerable sensitivity, which grows toward high energies
in a range available to accelerating laboratories.

{\it Experimental results.---}
The high-energy accelerators where laser Compton facilities have been operated 
for years, are listed on the upper energy scale of  Fig.~\ref{fig1}. 
As can be seen from the plot, 6~GeV storage rings (ESRF and VEPP) have low 
sensitivity while the higher energy colliders (HERA, SLC, LEP) have a great
potential for detecting the gravitational bending effect; see also Table~\ref{tab1} .
\begin{table}[htb]
\caption{\label{tab1} Sensitivity of different accelerators' Compton facilities to
the Earth's gravitational field.
}
\begin{ruledtabular}
\begin{tabular}{|c|c|c|c|c|c|}
Accelerator & Electron & Kinematic & $\omega_{max}$ & $\omega_{max}$ & Shift by \\
            & energy & factor      &  $n=1$         & $n=n_\oplus $  & gravity  \\
            &  GeV   & $x$         &  GeV          & GeV  & GeV  \\ \hline \hline
ESRF, VEPP  &  6.0   &   0.21    &   1.05  &     1.39  &   0.34  \\ \hline
HERA &  26.5  &    0.98  &     13.1  &     23.4  &    10.3  \\   \hline
SLC, LEP & 45.6  &     1.62  &     28.2   &    43.7  &    15.5 \\
\end{tabular}
\end{ruledtabular}
\end{table}
 Although all three machines are not operational anymore, one can analyze  
available data recorded by these accelerators where laser Compton setups were employed  
for polarimetry. Expected shifts of the maximal Compton energies are large
and so prominent that they would not have been missed if this magnitude gravitational influence was present there.
This is true for the HERA and SLC  but not for the LEP Compton polarimeter, which has
generated and registered many photons per machine pulse~\cite{LEP-polarimeter}. 
In this multi-photon regime, any shift of the Compton edge is convoluted with the 
laser-electron luminosity and can-not be disentangled and measured separately.

Unlike the LEP, the SLC polarimeter has operated in multi-electron mode and has analyzed 
the energies of interacted electrons using a magnetic 
spectrometer~\cite{slac-pol}. The spectrometer 
converted energies to positions, which then were detected by an array of  
Cherenkov counters.
The  position-energy correspondence has been derived from
the spectrometer magnetic field  strength according to the following expression:
\begin{equation}
S_x =\frac{296.45~GeV\cdot cm}{{\cal E}'} - 9.61~cm~,
\label{eqhorz}
\end{equation}
where $S_x$ is the position of the scattered electron with energy 
${\cal E}' = {\cal E}-\omega$. The scaling factor is quoted from  ref.~\cite{slac-pol}
and the offset, which depends on the electron beam position at the laser-electron interaction point, corresponds to a calibration from ref.~\cite{Gharibyan:2003fe}.
According to this  relation, the SLC polarimeter's Compton edge electrons with 17.4~GeV 
energy will enter the detector at a position of 7.43cm. This is what has been measured 
with $200\mu m$ statistical accuracy by a kinematic endpoint scan and is presented 
in Fig.~3-9 of  ref.~\cite{slac-pol}.
Could it happen that these authors have measured the GR-supported value of 1.9~GeV 
(at $n=n_\oplus$ from table~\ref{tab1}) instead of  17.4~GeV? 
Eq.~(\ref{eqhorz}) tells us that the 1.9~GeV electrons will have a position of 
146.4~cm  at the detector location, inconsistent with what has been measured (7.43cm).
Possible instrumental influence is limited to the initial electron beam position 
shift, less than 1~cm (to be contained in the accelerator's magnetic 
lattice~\cite{Chao:2013rba}) and 
an estimated accuracy of the magnetic spectrometer, better than 2\%. These factors 
add up to a maximum energy uncertainty or a possible offset of 1.4~GeV for the 
measured value of 17.4~GeV, reducing it to 16~GeV, which is still too high 
compared with the predicted value of 1.9~GeV.
I therefor conclude that the SLC polarimeter data do not support GR gravitational bending. 
   
\begin{figure}[h]
\centering
\includegraphics[scale=0.49]{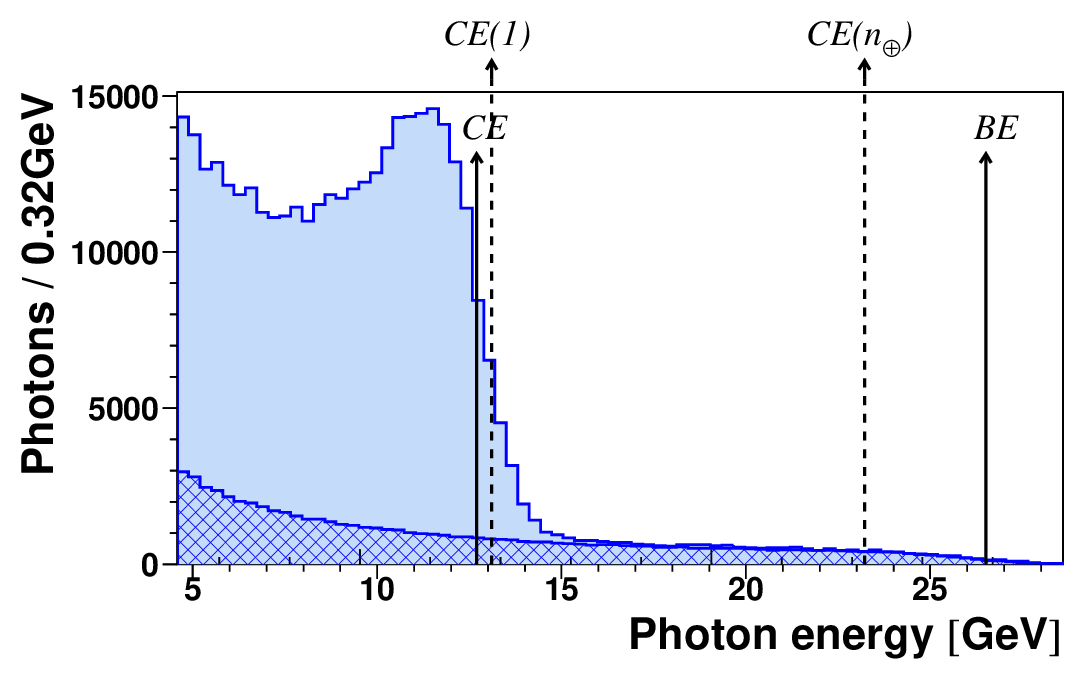}
\caption{\label{fig2}
HERA polarimeter Compton and Bremsstrahlung (hatched area) spectra.
Vertical solid lines show measured positions of  the Compton (CE) and Bremsstrahlung (BE) 
maximal energies. The dotted lines correspond to predicted Compton edge for
free (CE(1)) and Earth's gravitational (CE($n_\oplus$))  space.}
\end{figure}

At the HERA transverse polarimeter, Compton photons are registered by a calorimeter 
in single particle counting mode. A recorded Compton spectrum from ref.\cite{Barber:1992fc} 
is shown in Fig.~\ref{fig2} superimposed on a background  Bremsstrahlung distribution.
In contrast to the Compton scattering, in the Bremsstrahlung process the momentum 
transfer is not fixed, and any small $n\neq 1$ effect is smeared out and becomes 
negligible~\cite{Gharibyan:2003fe}.
Hence, following the analysis in  ref.~\cite{Gharibyan:2003fe}, I calibrate the energy 
scale according to the maximal Bremsstrahlung energy and show in Fig.~\ref{fig2}
the free space- and GR-predicted Compton edge energies (from table~\ref{tab1}), relative to 
the Bremsstrahlung edge. 
Comparing a measured maximal Compton energy of $12.7\pm0.1$~GeV from  
ref.~\cite{Gharibyan:2003fe} with the GR expectation of 23.4~GeV reveals a 
huge difference that can not be explained by any one of
the instrumental mis-measurement sources discussed in
 refs.~\cite{Barber:1992fc} and \cite{Gharibyan:2003fe} (or a total of
 these systematic uncertainties).
Therefore, I have to conclude that the HERA Compton experiment rules out
the GR prediction about gamma-ray bending.

From the SLC and HERA measurements and the derived or quoted numbers, it follows that both
Compton facilities have a much higher sensitivity to the Earth's gravitational refractivity 
than that of the GR value in Eq.~(\ref{refearth}), \hbox{$n_\oplus -1 = 1.39\times 10^{-9}$}. 
Indeed, as reported in ref.~\cite{Gharibyan:2003fe}, the anomalous  refractivity equals 
\hbox{$-(4.07\pm 0.05) \times 10^{-13}$} for the SLC 16.3--28.3~GeV photons and 
\hbox{$-(1.69\pm 0.47) \times 10^{-11}$} for the HERA 12.7~GeV gamma-rays.
At the time of the publication of  ref.~\cite{Gharibyan:2003fe} and up until now, the
source of this refractivity has remained unknown since possible contributions by a non-perfect
machine vacuum, electromagnetic stray fields, or hypothetical vacuum 
polarizations~\cite{Latorre:1995cv, Dittrich:1998fy, Scharnhorst:1998ix} 
are negligibly small ($< 10^{-20}$).  
Now, in light of real gravitational field interpretations, 
the observed bending ability of the laboratory vacuum could be attributed 
to Earth's gravity as the most influental and  likely source. 
Thus, combining the SLC and HERA results and multiplying  by a factor of 2
to obtain the integral bending,  
one can state that \hbox{12.7--28.3~GeV} gamma-rays are deflecting away from 
the Earth by \hbox{33.8--0.81$\times 10^{-12}$rad}. 

{\it Conclusions.---}
In order to test the gravitational deflection of photons at the Earth, I 
first described GR light bending in equivalent, optical refractivity terms.
Next, for the solution, I applied high-energy laser Compton scattering, 
which is extremely sensitive to any small refractivity due to its well-defined 
initial and final energy states and fixed momentum transfers.
Finally, I explored available experimental records from the SLC and HERA 
Compton polarimeters, finding with high confidence that gamma-rays are not 
bending according to GR. The observed energy dependence of gravitational bending 
is incompatible with a curved space approach and invalidates 
GR or other alternative,  purely geometrical, metric gravity 
theories described, for instance, in ref.~\cite{will-book}.

The SLC and HERA data also revealed a much smaller, negative 
deflection or repulsion of the high-energy photons from the Earth. 
Such a change of refractivity sign from positive(attractive) at low 
energies to 
negative(repelling) at high energies is a known feature of photon interactions 
with ordinary matter~\cite{xray-x-ray-refr-index}. 
This analogy could open new perspectives on quantum gravity, for which 
quantization of the GR geometrical gravitation is a major problem~\cite{Woodard:2009ns}.
With the detected energy dependent photon scattering,  gravity is
exposing an attribute belonging to momentum transfer interactions, 
which the  quantum approach can  more conventionally handle. 
A possible connection of the observed effects to the Planck scale, 
where quantum gravity should materialize, has been discussed previously  
in ref.~\cite{Gharibyan:2012gp}.
Nonetheless, despite its possible relation to the quantum regime,
repelling gravity is something quite unusual and
dedicated accelerator experiments are needed to check
the preliminary SLC and HERA  results on the negative refractivity.

Within the developed formalism of the laser Compton scattering in 
gravitational fields only the refraction of photon rays is taken into account.
The inclusion of a gravitational refractivity for high 
energy electrons can considerably reduce the magnitude of the described 
effects~\cite{priv1}. 
Then, the mentioned repelling gravity can be reinterpreted as a
decreased gravitational attraction for the photons compared to  
the electrons.

\end{document}